# III-V antiphase boundaries are not generated by Si or Ge substrate step edges

*C. Cornet[1,*], S. Pallikkara Chandrasekharan[1], A. Gilbert[2], M. Silvestre[2], R. Bernard[1], P. Turban[3], G. Patriarche[4], E. Tournié[2], L. Pedesseau[1] and J.-B. Rodriguez[2]*

[1] Univ Rennes, INSA Rennes, CNRS, Institut FOTON – UMR 6082, F-35000, France

[2] IES, University of Montpellier, CNRS, F- 34000, France

[3] Univ Rennes, CNRS, IPR (Institut de Physique de Rennes)–UMR 6251, F-35000 Rennes, France

[4] C2N, University of Paris Saclay, CNRS, UMR 9001, France

*Corresponding author. E-mail address: charles.cornet@insa-rennes.fr

**Abstract:**

We critically review recent experimental and theoretical advances on the valence- and structure-mismatched heteroepitaxy of III–V semiconductors on group-IV substrates. We examine key aspects including wetting behavior, atomic configurations at the hetero-interface, substrate passivation, and the stability of antiphase boundaries (APBs). By synthesizing these findings with pioneering studies, we propose a refined description of antiphase boundaries formation. Our analysis shows that hetero-interface formation is dictated by substrate terrace reconstruction, demonstrating that APBs are not generated at monoatomic step edges. This generalized III–V/IV growth framework provides new insights for the integration of III–V semiconductors on group-IV platforms and, more broadly, for valence-mismatched heteroepitaxy.

## 1. Introduction

Since the 1980s, the monolithic integration of III-V semiconductors on group-IV substrates (Si or Ge) has attracted significant research interest due to its potential for a wide range of applications. In active photonics, this integration has enabled the development of single photon sources,[1] GaAs/Si edge-emitting lasers,[2] integrated interband cascade lasers,[3] long-wavelength GaAs/Ge lasers,[4] and laser sources on monolithic InP/SOI platforms.[5] Beyond photonics, III-V/IV monolithic integration has also been explored for non-linear photonics, including GaP/Si microdisks [6] and V-groove nanopatterning,[7] as well as for high-performance electronic devices.[8] Another promising application of this integration is in multijunction photovoltaic solar cells, such as GaAsP/Si tandem structures,[9] InGaP/GaAs/Ge triple-junction architectures,[10] and InGaP/GaAs/Si configurations.[11] Additionally, this approach has been considered for photoelectrochemical water splitting applications, particularly for solar hydrogen production.[12,13]

However, the development of these devices has faced significant challenges, primarily due to the generation of numerous crystal defects during III-V/IV epitaxy. Among these defects, antiphase domains (APDs) have posed a major obstacle. To mitigate their propagation, researchers have progressively adopted the use of slightly misoriented substrates (also referred to as substrate miscut or offcut substrates). Miscut angles, typically ranging from 0.1° to 10°, have been shown to significantly improve layers quality.[14] However, the misorientation mismatches the natural cleavage planes of the crystal with those required for wafer processing and promotes step bunching or appearance of non-(001) facets during the growth, thereby complicating the fabrication of devices such as laser cavities.

While heteroepitaxy was already highly advanced by the early 1980s, most optoelectronic devices developed at that time relied on heterostructures composed of atoms sharing the same valence and crystal structure, such as III-V/III-V, II-VI/II-VI, or IV/IV systems. For example, in "valence-matched heteroepitaxy" of III-V semiconductors, only group III atoms (with a valence of $ns^2np^1$) and group V atoms (with a valence of $ns^2np^3$) are combined, as depicted in Fig. 1 (top left). This configuration ensures ideal charge sharing between atoms, regardless of the specific group III or V elements used. Additionally, this heteroepitaxy is "structure-matched", as the crystal structure of conventional semiconductors remains zinc-blende (nitrides are excluded from this discussion). At structure- and valence-matched III-V heterointerfaces, charge sharing remains continuous, and the primary growth challenges stem from lattice mismatches between the two III-V crystals and potential alloying fluctuations.

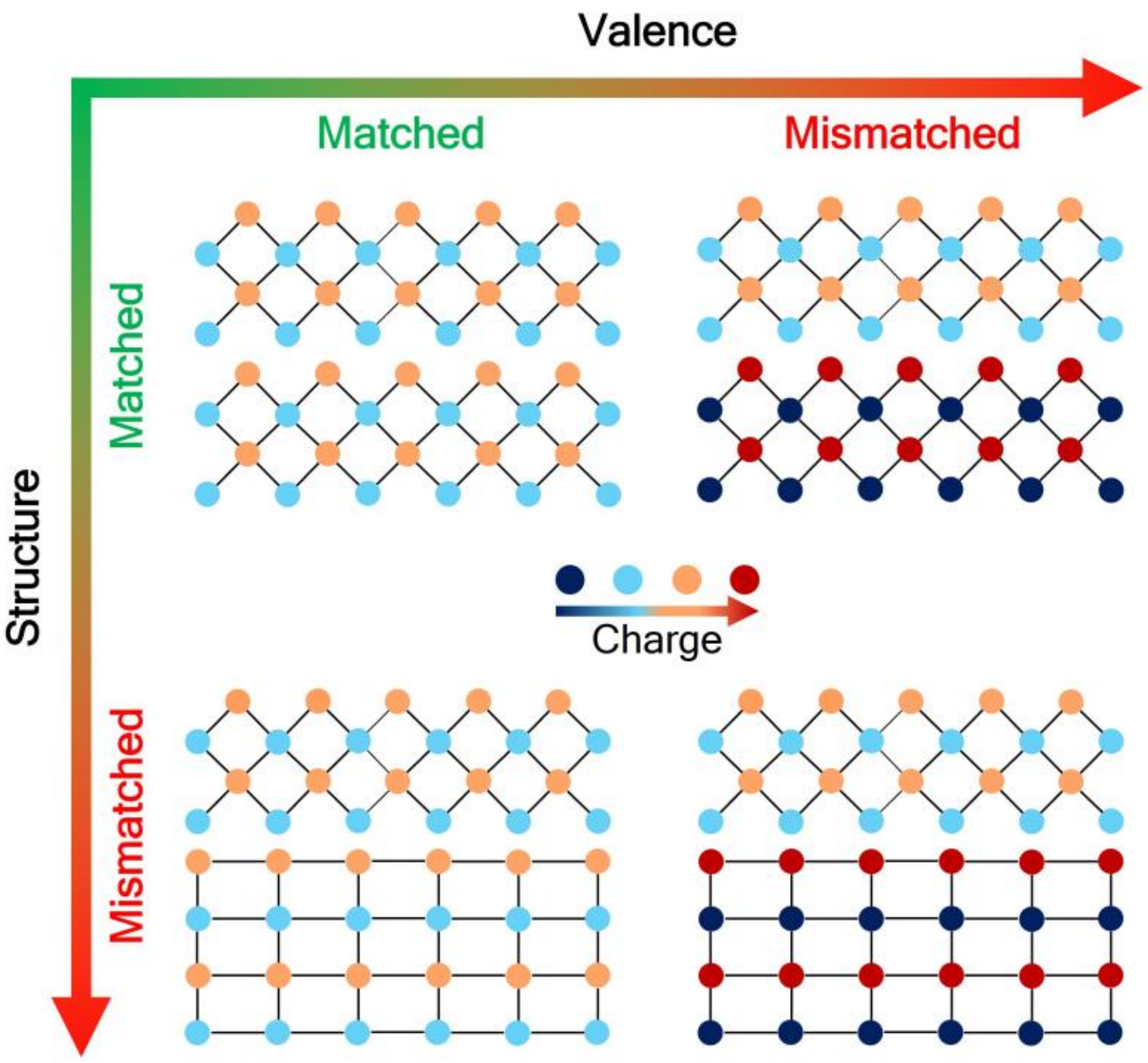

**Figure 1.** Proposed classification of hetero-epitaxial systems. Heteroepitaxy can be valence- and/or structure-(mis)matched, depending on the materials considered for hetero-integration.

In contrast, III-V/IV heteroepitaxy represents a valence-mismatched system, involving group III and V atoms (as described above) alongside group IV atoms (with a valence of $ns^2np^2$). The formation of a valence-mismatched hetero-interface introduces a charge discontinuity, which profoundly affects surface and interface energies, independently of lattice mismatch (Fig. 1, top right). This fundamentally alters the physics of growth, distinguishing it significantly from valence-matched heteroepitaxy. Furthermore, zinc-blende over diamond III-V/IV heteroepitaxy is also structure-mismatched (Fig. 1, bottom left), which can lead to the formation of antiphase or inversion domains due to the coexistence of different crystal orientations,[15] as shown in Fig. 2.

Thus, III-V/IV heteroepitaxy combines the challenges of valence-mismatched charge issues with those of structure-mismatched crystal systems (Fig. 1, bottom right). It is important to note that this dual mismatch is not universal: for instance, numbers of II-VI/III-V heteroepitaxial systems are valence-mismatched but structure-matched.

While Herbert Kroemer's seminal contributions laid the foundational understanding of the physical processes governing III-V/IV heteroepitaxy,[16,17] recent studies have revealed discrepancies with this initial framework.[18] In this work, we reexamine Kroemer's historical description by confronting it with the latest experimental and theoretical findings. Our analysis clarifies and revises key aspects of Kroemer's pioneering studies, particularly addressing the generation of antiphase boundaries (APBs) in III-V semiconductors. We demonstrate that APBs are not generated at the step edges of Si or Ge substrates, contrary to previous assumptions. Based on this insight, we propose a new scenario for APB formation, offering a refined understanding of the mechanisms underlying III-V/IV heteroepitaxy.

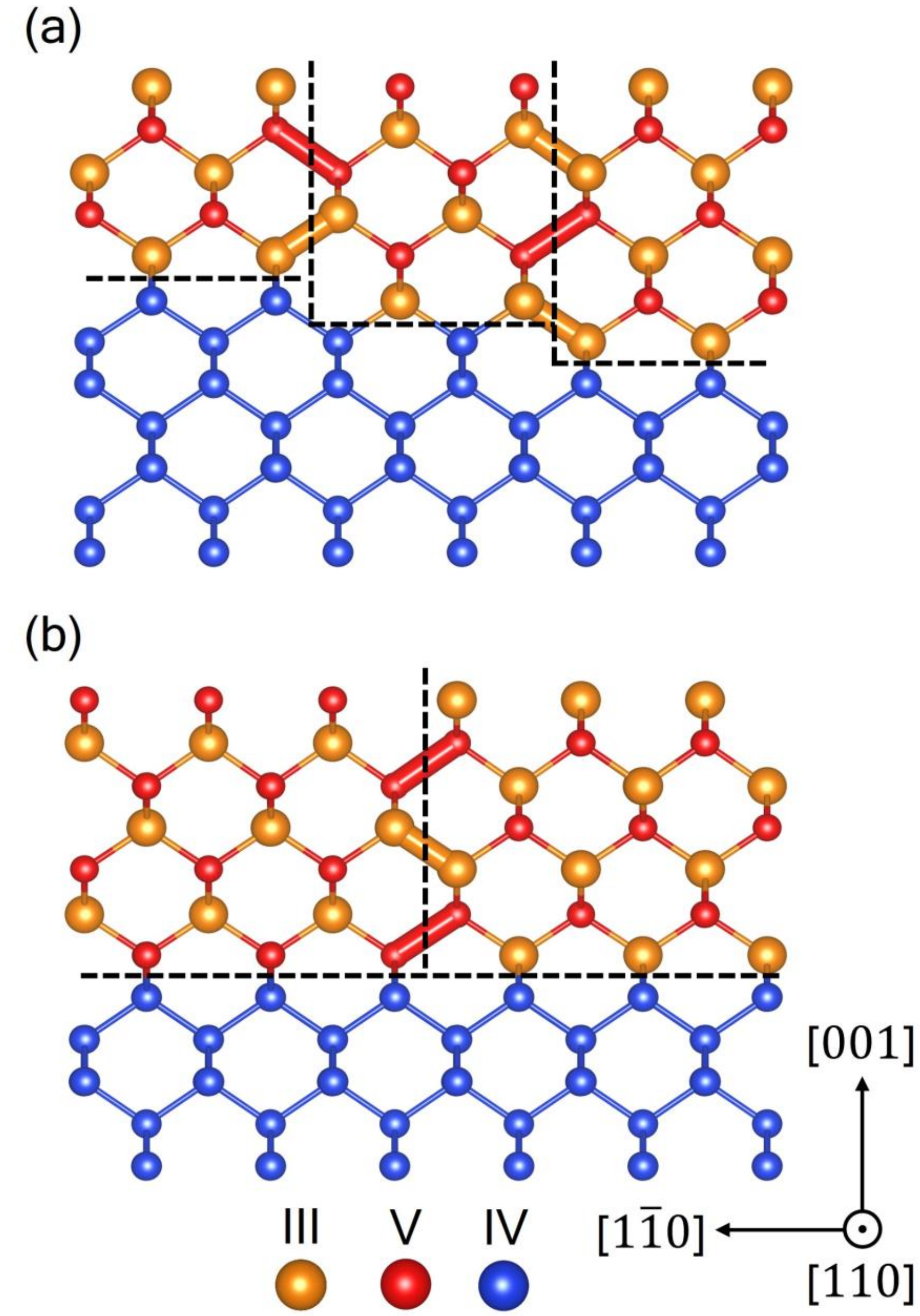

**Figure 2.** III-V AntiPhase Boundaries (APBs) and their connection to the group -IV substrate as reported in ref. [16,17]. APBs were considered as resulting from monoatomic steps at the group IV surface (a) or inhomogeneous coverage of a terrace (b).

## 2. Foundational studies on III-V/IV Hetero-interfaces: From substrate steps to antiphase boundaries

This section focuses on crystal defects (APBs) and atomic configurations (interfaces and substrate steps) specific to valence- and structure-mismatched heteroepitaxy. While the generation of misfit dislocations is not addressed here, we review the pioneering studies that have shaped our understanding of APBs and their relationship with substrate morphology.

APBs were first identified and discussed in the early 1980s in studies on III-V/IV heteroepitaxy, such as GaAs/Ge systems.[16] Their origin, formation, and stability were later explored in greater detail, notably in Kroemer's seminal work on GaAs/Si.[17] Kroemer proposed that monoatomic steps on the substrate surface invert the sublattice allocation of Ga and As atoms across the step, leading to the formation of APBs. This model assumes a perfectly abrupt interface between III-V atoms and the Si substrate, implying that APBs are inherently generated by group IV step edges, as illustrated in Fig. 2(a).

In the same study, Kroemer suggested using substrate miscut to promote the formation of double steps on the group IV surface, theoretically suppressing the generation of APBs. Subsequent research revealed the formation of 3D islands during the early stages of growth, attributed to strained Stranski–Krastanov nucleation.[19] Various strategies were proposed to overcome this 3D growth mode, sparking extensive research in this field with limited outcomes.

Since then, many studies in the literature have linked APBs to either monoatomic step edges on the substrate surface or, in the case of flat surfaces, to differential initial coverage of the Si surface by group-III or -V atoms (Figure 2(b)) .[15] These interpretations typically assume abrupt interfaces, as illustrated in Figure 1. Growth was often idealized as a 2D layer-by-layer process, with deviations and 3D growth attributed to strain effects.

Finally, the formation of double steps on the substrate surface has frequently been invoked to justify the use of substrate miscut. However, simultaneous direct evidence of step doubling and large-scale absence of APBs remains scarce in most studies, except in the work of Warren *et al*.[20] This gap is largely due to the challenges associated with characterizing buried interfaces and atomic crystal structures at large scales.[15,21,22]

However, this historical perspective was established while overlooking several studies that appear, at first glance, to be contradictory. In this context, the pioneering work of Petroff is particularly noteworthy.[23] Petroff not only reported a three-dimensional GaAs growth mode on Ge despite the near-lattice matching, but also proposed that APBs could originate from the coalescence of the islands, an interpretation in direct contradiction with Kroemer's viewpoint. This idea was later extended by Narayanan et al., [24] after the observation of a similar 3D growth mode in the quasi-lattice-matched GaP/Si system.[25] Concerning the atomic structure of the interface, Kroemer himself addressed in the same work the problem of valence-mismatched epitaxy, arguing that charge compensation must occur at the interface, thereby rendering abrupt interfaces inherently unstable.[17] This challenges by itself the prevailing model of APB formation based on step edges. In the following section, we examine recent findings that shed light on, and in part reconcile, these apparent contradictions in the literature in a generalized description.

## 3. Generalized III-V/IV Growth processes

### 3.1 Atomic structure of the III-V/IV Hetero-interface

Despite the recent progresses of electron microscopy, the 3D atomic structure of a complex hetero-interface cannot be resolved unambiguously, because the technique fundamentally averages the measured signal over atomic columns. In this context, Density Functional Theory (DFT) has been extensively used in recent studies to infer the relative stability of III–V/Si interfaces. The first detailed *ab initio* investigation of the GaP/Si interface was performed by Supplie *et al.* ,[26] who compared abrupt interfaces, where Si atoms are bonded exclusively to group-III or group-V atoms, with compensated interfaces, in which mixed atomic rows of Si and group-III/V atoms are present. Their calculations demonstrated that compensated interfaces, with compensation occurring within the very first monolayer, are the most stable over a wide range of chemical potentials. More recently, *ab initio* calculations of absolute interface energies for GaP/Si yielded values of 27.3 meV/Å$^2$ for the P-compensated interface and 23.4 meV/Å$^2$ for the Ga-compensated interface, both significantly lower than those of abrupt interfaces (typically varying between 70 and 30 meV/Å$^2$ over the whole chemical potential range) .[18,27]

Consistently, these theoretical predictions indicate that the most stable III–V/Si interface is one where half of the atoms in the terminating Si plane are replaced by group-III atoms. Experimental studies further corroborate this picture: both Metal-Organic Vapor Phase Epitaxy (MOVPE) growth of GaP/Si[28] and Molecular Beam Epitaxy (MBE) growth of GaAs/Si,[29] under terrace-driven conditions (*i.e.*, when III-V islands are smaller than terrace widths[30]), reveal a systematic correlation between APD formation and the terrace pattern of the Si substrate. Following the growth approach developed by Gilbert *et al.*, [29] a 50 nm -thick GaAs/Si sample was grown on a Si(001) substrate misoriented by 0.2° toward the [1-10] direction. At the end of the growth, an As amorphous capping layer was used to protect the surface, and further desorbed at 470+/-10 °C in the Scanning Tunneling Microscopy (STM) chamber for observation of the surface morphology. This procedure has been discussed and optimized in a previous work.[31] At a large scale of observation (Fig. 3(a)-1000*1000 nm²), the alternance of the domains at the surface of the GaAs has strong similarities with the expected step patterns at the Si surface, with a regular alternation of almost perfect straight lines, and kinked lines. This alternation, which periodicity fits perfectly with monoatomic steps with a 0.2° miscut, is almost identical to the one of $S_A$ and $S_B$ monoatomic steps at the Si surface.[32]

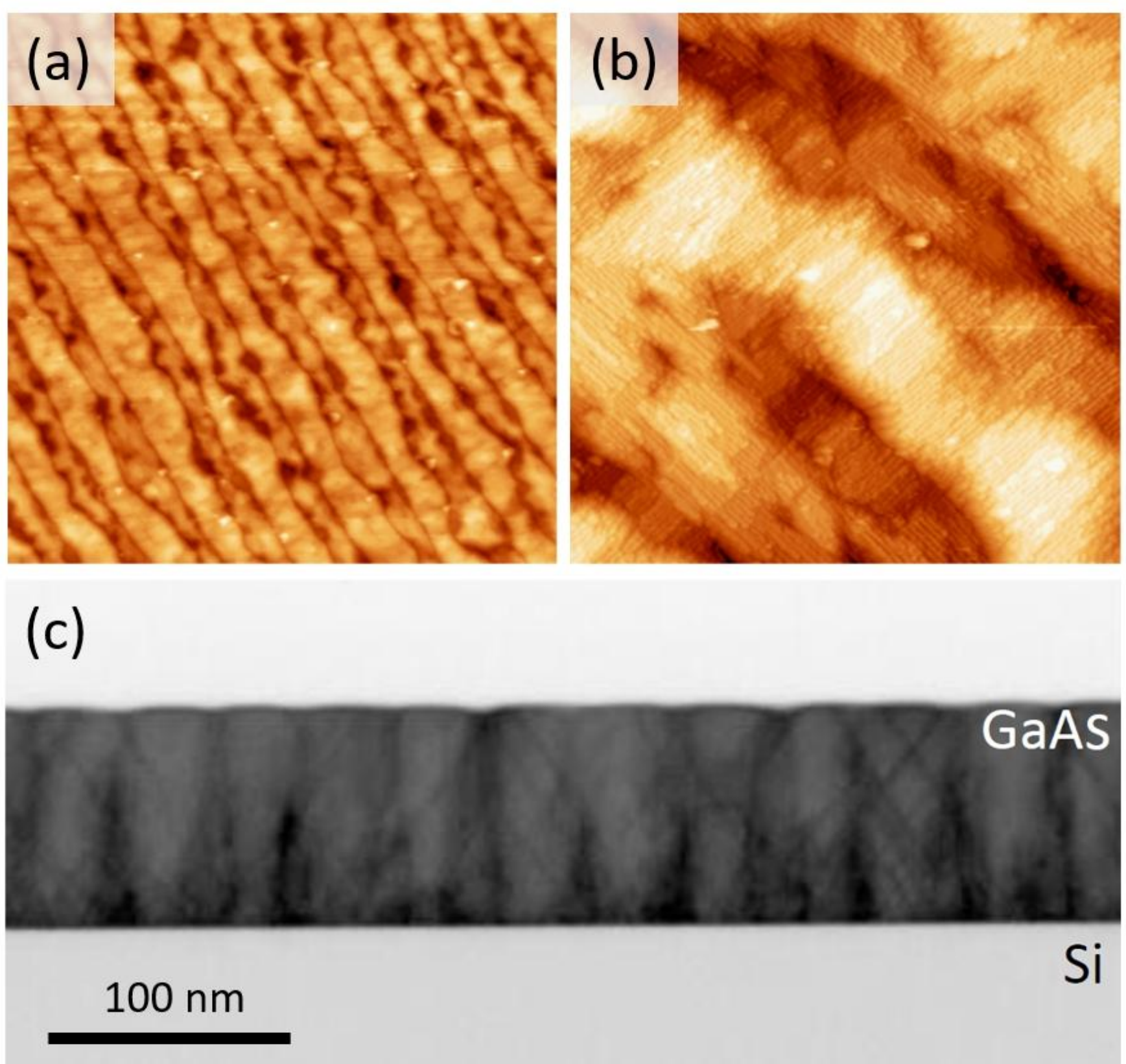

**Figure 3.** Typical Scanning Tunneling Microscopy (STM) images of a 50 nm-thick GaAs/Si(001-0.2°-off) sample on (a) a large area (1000*1000 nm²), and on (b) a smaller area (150*150 nm²). Transmission Electron Microscopy (TEM) cross-sectional image of a 100 nm-thick GaAs/Si(001-0.5°-off) sample.

Looking more closely at the surface (at the atomic scale), the situation is different. In fig. 3(b), a 150*150 nm² image of the same sample is provided. The APBs can be identified in the image when the surface reconstruction turns by 90°. At the atomic scale, the structure of APBs becomes different than the one expected for Si steps, with much less regularity of the frontiers between the steps and many apparent kinks everywhere. This can be understood considering that APB do not propagate in a straight way to the surface. This is illustrated with the cross-sectional TEM image of a thicker GaAs/Si sample provided in Fig. 3(c), where APDs are visible (as dark areas), revealing complex shapes in the material. The fluctuations of APBs propagating directions have been reported and discussed in many different works, for many different systems.[28–30,33,34] Consequently, it is absolutely not possible from these III-V/Si surface measurement to determine whether the III-V phases reproduce the Si step pattern or the Si terrace pattern. This will be discussed in next sections.

However, these results provide direct evidence that the orientation of surface dimers on the substrate fully dictates the polarity of the overgrown III–V material. In other words, the atomic arrangement of the III-V material over the Si terrace is not random but adopts a unique, well-defined interface configuration during the III-V nucleation on the substrate. This well-defined interface configuration is likely imposed by the need to equilibrate charges during the valence-mismatched heteroepitaxy.

Taken together, these experimental observations and theoretical insights converge on the conclusion that a single, stable atomic interface forms during III–V growth on Si terraces. Among the possible configurations, the group-III–compensated interface emerges as the most probable. Remarkably, this conclusion appears to hold universally, independent of growth method (MOVPE, MBE, or Hydride Vapour Phase Epitaxy - HVPE), growth regime (near or far from equilibrium), or the specific zinc-blende III–V material employed. Morever, the stable III- or V-compensated interfaces energies computed remain stable over all the chemical potential range, and thus, whatever the growth conditions considered. Consequently, attempts to alter the initial surface exposure to group-III or group-V species, or to implement alternate growth modes such as Migration-Enhanced Epitaxy (MEE) or Flow-rate Modulation Epitaxy (FME),[15,35] are not expected to modify the final atomic configuration of the interface. In summary, during the III-V/IV heteroepitaxial nucleation, the system invariably evolves kinetically toward a unique, most stable interface that is identical across the entire sample.

### 3.2 III-V Volmer-Weber 3D growth on group IV substrates

Building on the pioneering work of Petroff on GaAs/Ge[23] and of Ernst et al.[25] and Narayanan et al.[24] on GaP/Si, the initial stages of growth have been thoroughly analyzed in Ref. [18] and, more recently, in Refs. [36,37] for various III–V semiconductors grown on Si(001) and Si(111). Using both plan-view and cross-sectional microscopy, these studies confirmed that growth invariably starts with nm-scale 3D islands. Importantly, these islands are monodomain and grow laterally over distances much larger than the average terrace width of the substrate. No wetting layer was observed, and the role of stress on island evolution was found to be negligible. We note that these islands feature such a small size and high density that they cannot typically be resolved by conventional atomic force microscopy, while their rapid coalescence can mimic a continuous film in cross-sectional TEM, leading to their historical misidentification as 2D layers.

The question of wetting was further investigated using DFT, where absolute surface and interface energies as well as spreading parameters were calculated under varying chemical potentials.[18,27,38] For GaP/Si, these studies demonstrated that the system follows a Volmer–Weber growth mode at thermodynamic equilibrium, a result likely generalizable to other III–V semiconductors grown on Si or Ge, given the comparable orders of magnitude. Interestingly, the III–V/Si hetero-interface energy was shown to be lower than initially expected when the interface is charge-compensated, underlining the crucial role of surface passivation.

In fact, the surface energy of a bare Si surface is relatively high,[18] making it highly reactive toward adsorbing species such as oxygen, hydrogen, III or V atoms. Since oxygen and hydrogen are typically removed during chemical or thermal substrate preparation, group-III or group-V atoms present in the growth chamber can act as an efficient passivation monoatomic layer, significantly reducing the Si surface energy and thereby favoring the Volmer–Weber growth mode.[38] This conclusion is robust and independent of the growth technique employed (MOVPE, MBE, or HVPE), although minor differences may occur for hydrogenated surfaces without affecting the overall surface or interface energy hierarchy. While these findings were initially established for III–V/Si, a similar behavior is expected for III–V growth on Ge. A comprehensive summary of the relevant surface and interface energies for the III–V/Si system is provided in Ref. [38].

**3.3 Antiphase Boundaries formation**

The last missing piece in this framework is to explain how III–V islands can grow monodomain (*i.e.*, without forming APBs) across monoatomic substrate steps, as observed experimentally.[18] This observation is essential to understanding the influence of substrate miscut on APD distribution during growth, particularly in the nucleation-driven regime (*i.e.* when islands are larger than terraces widths[30]).

This issue was recently addressed by Gupta *et al.* [39], who demonstrated using DFT that it is energetically far more favorable to adapt the silicon step edge through a change in interface configuration than to form an APB. APBs are thus metastable defects formed in non-equilibrium conditions which explains why they are strongly impacted by temperature evolution or annealing during the growth.[40,41] This is fundamentally due to the fact that both group-III- and group-V-compensated interfaces are stable, whereas APB formation carries a much higher energy cost. Consequently, APBs only form when no alternative is available, namely, during the coalescence of independent 3D islands. In a *gedanken* experiment, a near-to-equilibrium single III–V island growing on a group-IV substrate would thus remain monodomain over all the wafer, even after crossing an infinite number of monoatomic steps.

Based on these considerations, a realistic picture of APB formation can now be drawn (Fig. 4). At the onset of growth (Fig. 4a), 3D islands nucleate, each stabilized by a group-III–compensated interface on the underlying substrate terrace. While step edges are often considered preferential nucleation sites, such as in the Burton-Cabrera-Frank (BCF) crystal growth kinetics approach,[42] our analysis suggests that the precise location of initial nucleation (whether at steps or on terraces) does not fundamentally dictate the final film polarity. Because initial nuclei rapidly expand to cover multiple substrate atoms, the number of substrate surface terrace atoms involved quickly outweighs the number of substrate surface step atoms. Consequently, the selection of polarity is predominantly driven by the minimization of the hetero-interface energy to form the most stable planar configuration, a mechanism particularly critical in valence-mismatched epitaxy where interface energetics play a dominant role. Consequently, Figure 4 does not explicitly specify the initial nucleation site.

For illustration, (111) faceting is assumed in Fig. 4a, although other facets may appear depending on the material system, strain and growth method which can have an impact on the shape of the islands. Between islands, the surface is passivated by a single atomic layer of group-III or group-V atoms. The nature of this passivation depends on the chemical potential (*i.e.*, the V/III ratio),[38] but this does not fundamentally alter the process. Here, a group-V passivation is assumed to reflect the V-rich conditions typically used in epitaxy. In the second stage (Fig. 4b), islands expand in all directions, with growth rates dictated by the exposed facets and the specific III–V material. As shown, an island can overgrow a monoatomic step while remaining monodomain by adapting its interface from a group-III– to a group-V–compensated configuration (or vice versa). In the third stage (Fig. 4c), neighboring islands coalesce. An APB forms at the junction where the two distinct interfacial configurations meet, a process ultimately induced by the propagation of monodomain islands over monoatomic step edges. In Fig. 4c, for the sake of simplicity, a vertical APB is considered, corresponding to the situation where the two domains grow at the same rate.[30] However, the burying of one phase by another can lead to kinked or inclined APBs that follow either the facets of the initial islands, the facets bounding the newly formed APB or other intermediate directions.

This three-step scenario provides the only viable pathway for APB formation during III–V/IV valence- and structure-mismatched heteroepitaxy and forms the basis of the generalized growth framework that reconciles experimental observations.

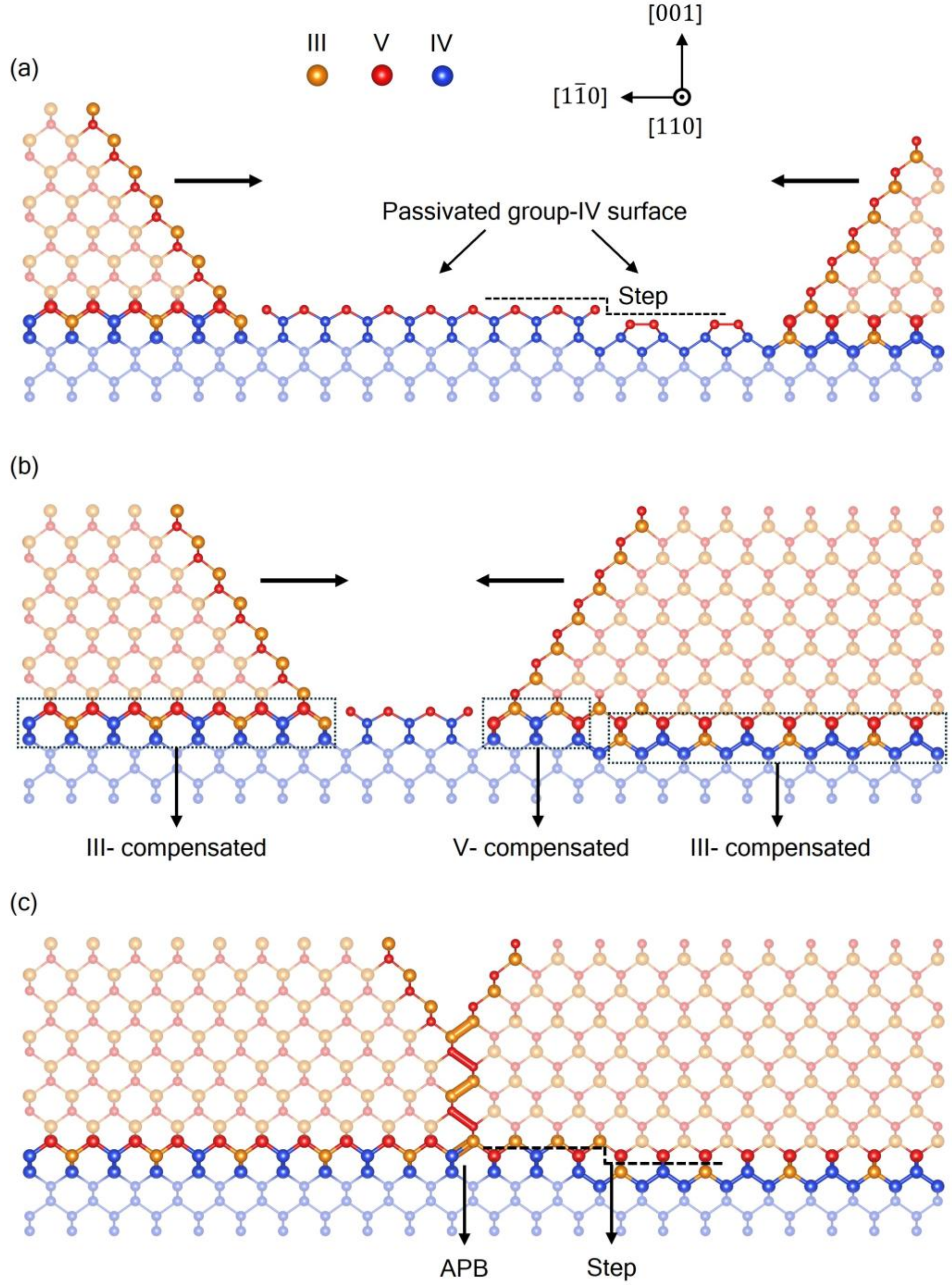

**Figure 4.** Generalized mechanism of III-V antiphase boundary (APB) formation on group IV substrate proposed in this work. Atoms belonging to bulk materials are made transparent for clarity. (a) Growth initiates with 3D III–V islands, each stabilized by the same compensated III–V/IV interface; the island polarity is therefore dictated by the substrate terrace orientation. The group-IV surface between islands is passivated by group-V or group-III atoms. (b) Islands expand laterally across terraces and over monoatomic steps while preserving a single domain, by adapting their interface configuration at the steps. (c) APBs arise upon the coalescence of islands nucleated on adjacent terraces with different interfacial configurations.

## 4. Discussion

The role of substrate miscut during structure-mismatched III–V/IV heteroepitaxial growth and burying of antiphase domains have been previously clarified[30] and experimentally investigated in both nucleation-driven (III-V islands larger than terrace width)[43] and terrace-driven (III-V islands smaller than terrace width)[29,44] regimes. Essentially, the high surface symmetry of the group IV substrate impedes single-phase bulk III-V epitaxy. The miscut, however, breaks this surface symmetry by suppressing the formation of atomic steps in one of the two main crystallographic directions, thereby enabling the progressive recovery of the III-V material's intrinsic surface symmetry. This is achieved through growth kinetics by inducing anisotropy in incorporation rates or promoting directional surface adatom diffusion, which facilitates APD burying.

Here, we emphasize the advantage of using a miscut to engineer the group-IV substrate steps prior to growth. At first glance, the APB formation mechanism described above does not appear radically different from that proposed by Kroemer.[17] In principle, a perfectly double-stepped surface would yield an APB-free III–V epilayer. However, this ideal situation is rarely achieved at typical growth temperatures, and APDs are most of the time observed at the onset of growth, even in high-quality III–V/Si samples.[15,21] This indicates the presence of significant minority terraces on the Si substrate before III–V deposition. Moreover, strong spatial inhomogeneities across the wafer have been reported, attributed to local temperature variations during or prior to growth.[45] Therefore, instead of targeting a perfect double-step surface, which is difficult to verify and control *in situ*, the previously described APB formation mechanism suggests a degree of freedom: as long as one type of group-IV terrace is sufficiently small, small APDs are generated and rapidly buried. This condition is easily met for wafers with large miscut angles but is more challenging for low-miscut wafers.[40] It also shows the great interest to promote steps aligned unidirectionally as much as possible during the group-IV wafer preparation stage.

This framework also clarifies how III–V growth conditions, which control island size, strongly influence APD distribution. Especially, while an initial low-temperature III-V growth step promotes a higher density of smaller islands and enhances the spatial correlation between overlayer APBs and substrate atomic steps, higher temperatures conversely tend to decorrelate these features. Consequently, careful preparation of the group-IV substrate surface prior to III–V growth is critical for achieving high-quality epitaxial layers. Large-scale monolithic integration of III–V semiconductors on low-miscut (<1°) group-IV wafers therefore requires substrate manufacturers to provide accurate miscut angles and directions, and the development of processes by research teams to ensure chemical or thermal pre-treatment that optimizes step arrangement and wafer-scale homogeneity. More specifically, a well-ordered Si step morphology can be achieved through simple annealing at temperatures typically above 900 °C with thick wafers. However, this comes at the cost of a high thermal budget and increased contamination. Conversely, more complex approaches, such as Si epitaxy or heating under plasma exposure,[46] can achieve the same ordered atomic steps with a cleaner surface at significantly lower temperatures. A similar trend applies to the Ge surface, albeit at much lower temperatures. Finally, we emphasize that an effective approach to bypass the step issue on the Si(001) surface is to intentionally promote the formation of stable {111} facets, rather than maintaining the natural stepped (001) surface. This has been shown to significantly reduce the density of generated antiphase boundaries.[47]

## 5. Conclusion

In this work, we have carefully reviewed recent experimental and theoretical advances on the epitaxy of III–V semiconductors on group-IV substrates. From this analysis, we propose a refined description of antiphase boundary (APB) formation, demonstrating that APBs cannot be generated at monoatomic step edges. Instead, III–V/IV valence-mismatched heteroepitaxy promotes the preferential formation of well-defined, stable compensated hetero-interfaces, which ultimately determine the polarity of the growing crystal. This framework highlights the central role of substrate terrace reconstruction in governing hetero-interface formation and APB generation. By establishing a generalized picture of III–V/IV growth processes, our work provides new insights not only for the integration of III–V semiconductors on group-IV platforms, but also for the broader understanding of valence-mismatched heteroepitaxy.

**Acknowledgments:**

This research was supported by the French National Research NUAGES (Grant No. ANR-21-CE24-0006), the EQUIPEX+ NANOFUTUR (Grant No. ANR-21-ESRE-0012), HYBAT (Grant No. ANR-21-ESRE-0026), and PIANIST (Grant No. ANR-21-CE09-0020) projects.

**Declaration of generative AI and AI-assisted technologies in the writing process**

During the preparation of this work the author(s) used Le Chat/Mistral AI to improve the readability of the article. After using this tool/service, the author(s) reviewed and edited the content as needed and take(s) full responsibility for the content of the publication.

**Author Contributions:**

The manuscript was written through contributions of all authors. All authors have given approval to the final version of the manuscript.

## Table of Contents Graphic

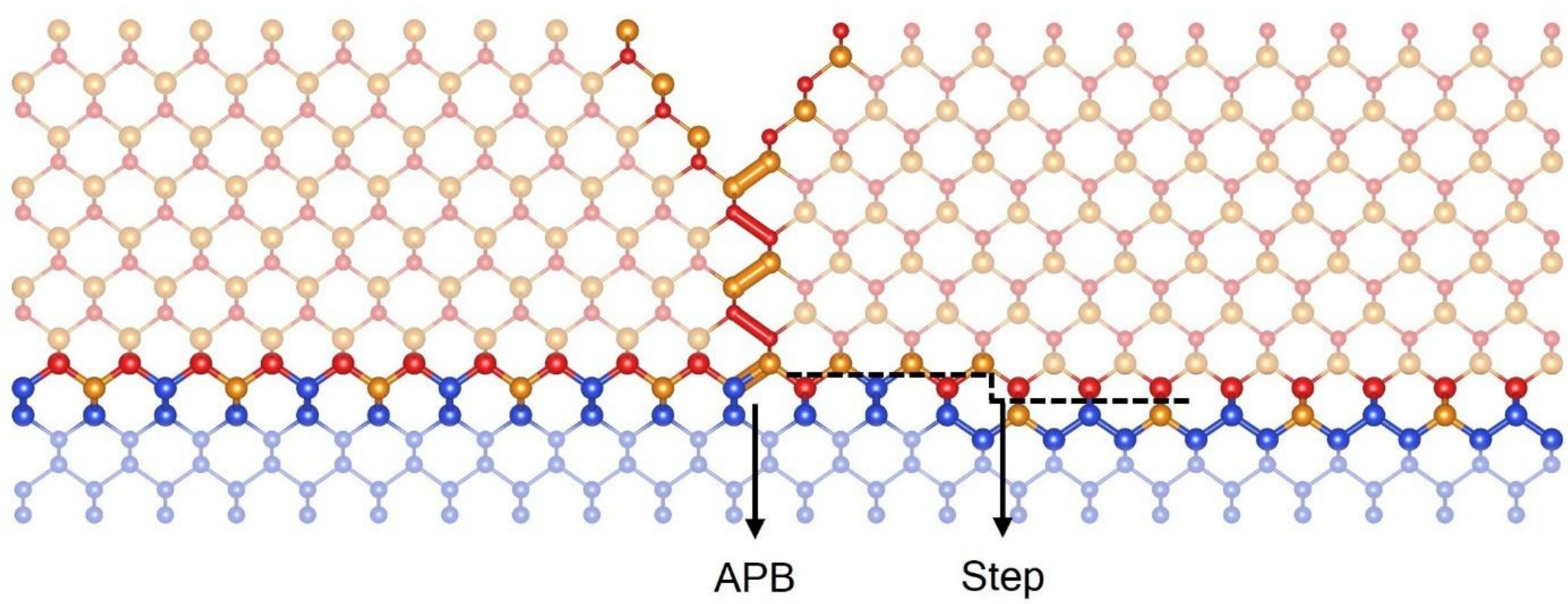

This work synthesizes recent advances in III–V/IV heteroepitaxy, analyzing wetting, atomic interfaces configuration, passivation, and Anti-phase Boundaries (APB) stability. It refines APB formation understanding, showing substrate terrace reconstruction—not monoatomic steps—dictates APB formation. This framework offers new insights for III–V/IV integration and valence-mismatched heteroepitaxy.